\documentclass[twocolumn,english,pra, aps, twocolumn,superscriptaddress]{revtex4-1}
\usepackage[T1]{fontenc}
\usepackage[latin9]{inputenc}
\usepackage{amsmath}
\usepackage{graphicx}

\makeatletter
\usepackage{xcolor}
\usepackage[colorlinks,citecolor=blue,linkcolor=blue]{hyperref}

\makeatother

\usepackage{babel}
\begin{document}
\title{Optimizing Thermodynamic Cycles with Two Finite-Sized Reservoirs}
\author{Hong Yuan}
\address{Graduate School of China Academy of Engineering Physics, No. 10 Xibeiwang
East Road, Haidian District, Beijing, 100193, China}
\author{Yu-Han Ma}
\email{yhma@gscaep.ac.cn}

\address{Graduate School of China Academy of Engineering Physics, No. 10 Xibeiwang
East Road, Haidian District, Beijing, 100193, China}
\author{C. P. Sun}
\email{suncp@gscaep.ac.cn}

\address{Graduate School of China Academy of Engineering Physics, No. 10 Xibeiwang
East Road, Haidian District, Beijing, 100193, China}
\address{Beijing Computational Science Research Center, Beijing 100193, China}
\begin{abstract}
We study the non-equilibrium thermodynamics of a heat engine operating
between two finite-sized reservoirs with well-defined temperatures.
Within the linear response regime, it is found that the uniform temperature
of the two reservoirs at final time $\tau$ is bounded from below
by the entropy production $\sigma_{\mathrm{min}}\propto1/\tau$. We
discover a general power-efficiency trade-off depending on the ratio
of heat capacities ($\gamma$) of the reservoirs for the engine. And
a universal efficiency at maximum average power of the engine for
arbitrary $\gamma$ is obtained. For practical purposes, the operation
protocol of an ideal gas heat engine to achieve the optimal performance
associated with $\sigma_{\mathrm{min}}$ is demonstrated. Our findings
can be used to develop an general optimization scenario for thermodynamic
cycles with finite-sized reservoirs in real-world circumstances.
\end{abstract}
\maketitle
\emph{Introduction}.--The thermodynamic constraints exist in all
kinds of energy-conversion machines. Among these constraints, Carnot
efficiency serves as the upper bound for efficiency of heat engines.
Such a bound is only achieved by reversible thermodynamic cycles under
quasi-static limit \citep{Carnoteff}, and is therefore not tight
for practical heat engines with finite cycle time. Considering the
restriction of operation time, abundant tighter thermodynamic constraints
were obtained for finite-time thermodynamic cycles \citep{andresen1984thermodynamics,StochasticThermodynamics,Holubec2017,kosloff2019quantum,tu2021abstract}.
For example, efficiency at maximum power (EMP) \citep{yvon1955saclay,chambadal1957recuperation,novikov1958efficiency,CA,BroeckPRL2005,IzumidaEPLStochatic,schmiedl2008efficiency,Tu2008JPhysAMathTheor41_312003,EspositoPRL2010,Wang2012,2013EfficiencyBroeck},
trade-off relation between power and efficiency \citep{2015Efficiency,tradeoffholubec,long2016efficiency,TradeoffrelationShiraishi,CavinaPRLtradeoffrelation,Constraintrelationyhma},
and thermodynamic uncertainty relation (TUR) \citep{barato2015thermodynamic,horowitz2017proof}.
In particular, the power-efficiency trade-off determines the feasible
operation regime for finite-time heat engines and has attracted considerable
attention.

Recently, to deal with another practicality that the heat is basically
stored by a finite amount of material with finite heat capacity, the
finiteness of the reservoir size is also taken into account as a physical
restriction on thermodynamic cycles \citep{ondrechen1981maximum,ondrechen1983generalized,leff1987available,izumida2014work,wang2014optimization,johal2016optimal,tajima2017finite,2020-finite-size}.
This issue is crucial for responding to the increasingly severe energy
crisis with limited material resources. And the efficiency at maximum
work (EMW) \citep{ondrechen1981maximum,leff1987available,johal2016near,2020-finite-size}
and efficiency at maximum average power (EMAP) \citep{ondrechen1983generalized,izumida2014work,wang2014optimization,2020-finite-size}
were proposed as typical thermodynamic constraints in this case.

As two fundamental restrictions in energy conversion processes, the
finiteness of operation time and reservoir size usually coexist in
real-world circumstances. Hence, a more practical question naturally
arises: \textit{Is there a power-efficiency trade-off associated with
finite-sized reservoirs}? In this Letter, we address this question
by studying the finite-time performance of a linear irreversible heat
engine operating between two finite-sized reservoirs. We discover
a general trade-off relation between power and efficiency. And a universal
EMAP is obtained. Furthermore, we find the optimal operation of the
engine to achieve the boundary of the trade-off.
\begin{figure}
\includegraphics[width=8.5cm]{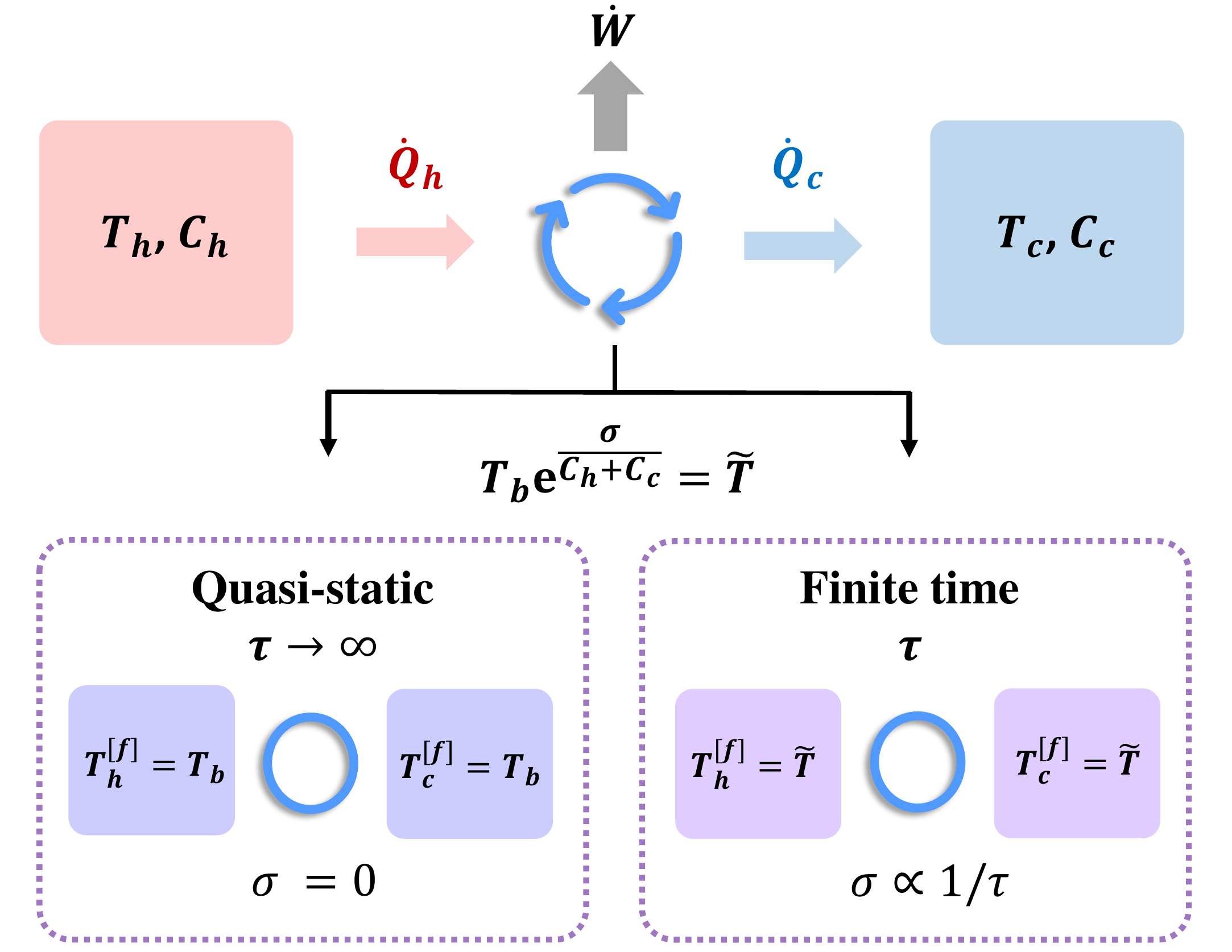}\caption{\label{fig:Demonstration-of-the}Demonstration of a heat engine operating
between two finite-sized heat reservoirs. The heat engine operates
between a finite-sized hot (cold) reservoir with initial temperature
$T_{h}^{[i]}$ ($T_{c}^{[i]}$). The heat engine stops working when
the two reservoirs reach a the final uniform temperature $T_{c}^{[f]}=T_{h}^{[f]}\equiv\tilde{T}$.
$C_{c}$ ($C_{h}$ ) denotes the heat capacity of the cold (hot) reservoir.
The increase in entropy production $\sigma$ will higher the final
temperature $\tilde{T}$ {[}Eq. (\ref{eqfinal T}){]}.}
\end{figure}

\emph{The minimum entropy production and the uniform temperature.}--
As illustrated in Fig. \ref{fig:Demonstration-of-the}, we consider
a linear irreversible heat engine operating between a hot reservoir
with initial temperature $T_{h}^{[i]}$ and a cold reservoir with
initial temperature $T_{c}^{[i]}$. Both of these two reservoirs are
of finite size with the heat capacity $C_{h}$ and $C_{c}$, respectively.
As follows, we focus on the case of constant heat capacity $C_{h(c)}$.
From the initial time $t=0$, the engine converts the heat to work
consecutively through a control parameter $\lambda$ until the two
reservoirs finally reach the thermal equilibrium state at $t=\tau$
with a uniform temperature $T_{c}^{[f]}=T_{h}^{[f]}\equiv\tilde{T}$.
Here, we stress that the heat capacity of at least one reservoir needs
to be finite, otherwise the temperature of the two reservoirs will
always maintain their initial values instead of reaching the same
within finite time. In the following, we adopt the assumptions used
in the Refs. \citep{izumida2014work,wang2014optimization,2020-finite-size}:
\textbf{(i)} both of the two reservoirs relax rapidly such that they
are always in the quasi-equilibrium states with time-dependent temperatures
$T_{h}(t)$ and $T_{c}(t)$;\textbf{ (ii)} the total operation time
$\tau$ (macro time scale) is much larger than the cycle time $\tau_{\mathrm{c}}$
(micro time scale, treat as a unit of time hereafter), and hence the
engine undergoes sufficiently many cycles, namely, $M\equiv\tau/\tau_{c}\gg1$,
before it stops operating.

The entropy production rate reads $\dot{\sigma}=-\dot{Q}_{h}/T_{h}+\dot{Q}_{c}/T_{c}$,
where $\dot{Q}_{h}=-C_{h}\dot{T}_{h}$ represents the heat absorption
from the hot reservoir to the engine of a cycle, and $\dot{Q}_{c}=C_{c}\dot{T}_{c}$
is the heat release from the engine to the cold reservoir of a cycle.
As a result, the total entropy production $\sigma(\tau)\equiv\int_{0}^{\tau}\dot{\sigma}dt$
is

\begin{equation}
\sigma(\tau)=C_{c}\ln\frac{\tilde{T}}{T_{c}^{[i]}}+C_{h}\ln\frac{\tilde{T}}{T_{h}^{[i]}}.\label{eq:entropy-production}
\end{equation}
The uniform temperature $\tilde{T}$ is thus determined by the entropy
production as

\begin{equation}
\tilde{T}=\tilde{T}(\sigma)=\left[T_{h}^{[i]}\right]^{\frac{1}{\gamma+1}}\left[T_{c}^{[i]}\right]^{\frac{\gamma}{\gamma+1}}\exp\left[\frac{\sigma}{C_{h}+C_{c}}\right],\label{eqfinal T}
\end{equation}
namely the uniform temperature rises as the entropy production increases.
Here the heat capacity ratio $\gamma\equiv C_{c}/C_{h}$ quantifies
the asymmetry in size of the reservoirs. $T_{b}\equiv\tilde{T}(\sigma=0)$
is the final temperature in the reversible case with no entropy production.
The reversible case is discussed in the Supplementary Materials (SM)
\citep{S-M}.

Then, we exploit the linear irreversible thermodynamics to obtain
$\sigma(\tau)$ as well as $\tilde{T}$ explicitly in the finite-time
regime. Under the tight-coupling condition $q\equiv L_{21}/\sqrt{L_{11}L_{22}}$=1,
the entropy production rate reads \citep{izumida2014work,wang2014optimization}

\begin{equation}
\dot{\sigma}=\frac{\dot{Q}_{h}^{2}}{L_{22}}=\frac{C_{h}^{2}\dot{T}_{h}^{2}}{L_{22}},
\end{equation}
where $L_{ij}$ ($i,j=1,2$) is the Onsager coefficient, and $L_{22}$
corresponds to the thermal conductivity \citep{izumida2009onsager,proesmans2015onsager,izumida2021hierarchical}.
The adopted tight-coupling condition can be practical realized, e.g.,
by a finite-time ideal gas Carnot engine \citep{izumida2009onsager}.

The Cauchy-Schwarz (C-S) inequality

\begin{equation}
\left[\int_{0}^{\tau}\left(\sqrt{\dot{\sigma}}\right)^{2}dt\right]\left(\int_{0}^{\tau}dt\right)\geq\left(\int_{0}^{\tau}\sqrt{\dot{\sigma}}dt\right)^{2}\label{eq:CS-in}
\end{equation}
implies that the entropy production $\sigma(\tau)=\int_{0}^{\tau}\dot{\sigma}dt=\int_{0}^{\tau}(\sqrt{\dot{\sigma}})^{2}dt$
has a lower bound, namely, \citep{S-M}

\begin{equation}
\sigma(\tau)\geq\frac{\Sigma_{\mathrm{min}}}{\tau}\equiv\sigma_{\mathrm{min}}.\label{eq:sigma min}
\end{equation}
In this inequality, only the first order of $\tau^{-1}$ is kept in
the long-time regime \citep{Wang2012}, and the equal sign is saturated
with constant entropy production rate, i.e., $\dot{\sigma}=\Sigma_{\mathrm{min}}/\tau^{2}$
($\dot{Q}_{h}=\sqrt{L_{22}\Sigma_{\mathrm{min}}}/\tau$). The minimum
dissipation coefficient $\Sigma_{\mathrm{min}}\equiv(\int_{T_{h}^{[i]}}^{T_{b}}C_{h}dT_{h}/\sqrt{L_{22}})^{2}$,
characterizing how irreversible entropy production increases away
from the reversible regime, is a $\tau$-independent dissipation coefficient.
Generally, $\Sigma_{\mathrm{min}}$ depends on the specific form of
$L_{22}$ and relates to the thermodynamic length \citep{Ruppeiner1979,salamon1983thermodynamic,Crooks2007,izumida2021hierarchical}.
In the simplest case with constant $L_{22}$, $\Sigma_{\mathrm{min}}=C_{h}^{2}[T_{h}^{[i]}-T_{b}]^{2}/L_{22}$.
The typical $1/\tau$-scaling of irreversibility shown in Eq. (\ref{eq:sigma min})
has also been discovered in the finite-time isothermal processes \citep{SekimotoJPSJ,schmiedl2008efficiency,Constraintrelationyhma,yhmaoptimalcontrol,2020IEGyhma}.

We remark here that although the minimum entropy production $\sigma_{\mathrm{min}}$
in Eq. (\ref{eq:sigma min}) is obtained with the tight-coupling condition,
$\sigma_{\mathrm{min}}$ actually serves as the overall lower bound
for entropy production $\sigma$ with arbitrary $q$. This is because
$\sigma$ decreases monotonically with the increase of $\left|q\right|$
(See SM \citep{S-M} for strict proof). Therefore, for general cases
within the linear response regime, the uniform temperature is bounded
from below by the minimal entropy production as $\tilde{T}\geq\tilde{T}(\sigma_{\mathrm{min}})$. 

\emph{Trade-off between power and efficiency.} -- The work output
in the whole process is $W(\tau)=Q_{h}(\tau)-Q_{c}(\tau)$, where
$Q_{h}(\tau)=C_{h}(T_{h}^{[i]}-\tilde{T})$ and $Q_{c}(\tau)=C_{c}(\tilde{T}-T_{c}^{[i]})$.
The maximum extractable work $W_{\mathrm{max}}\equiv\mathrm{lim_{\sigma\rightarrow0}}W(\tau)$
\citep{S-M} is achieved in the reversible case. Note that $W(\tau)$
is a monotonically decreasing function of $\tilde{T}$ \citep{S-M},
which indicates that, referring to Eq. (\ref{eqfinal T}), the entropy
production will reduce $W(\tau)$ in comparison with $W_{\mathrm{max}}$.
In this sense, we define the finite-time dissipative work 

\begin{equation}
W_{d}\equiv W_{\mathrm{max}}-W(\tau)=\left(C_{h}+C_{c}\right)\left(\tilde{T}-T_{b}\right).\label{eq:Wd}
\end{equation}
It follows from Eqs. (\ref{eqfinal T}), (\ref{eq:sigma min}), and
(\ref{eq:Wd}) that the constraint on dissipative work is explicitly
obtained as $W_{d}\geq T_{b}\Sigma_{\mathrm{min}}/\tau\equiv W_{d}^{(\mathrm{min})}$
\citep{S-M}. In terms of $W_{d}$, the efficiency in the finite-time
case, $\eta\equiv W(\tau)/Q_{h}(\tau)$, reads

\begin{equation}
\eta=\frac{W_{\mathrm{max}}-W_{d}}{W_{\mathrm{max}}/\eta_{\mathrm{MW}}-W_{d}/(1+\gamma)},\label{eq:eta-tau}
\end{equation}
where the efficiency at maximum work (EMW) \citep{S-M}

\begin{equation}
\eta_{\mathrm{MW}}\equiv1-\gamma\left[\frac{\eta_{\mathrm{C}}}{1-\left(1-\eta_{\mathrm{C}}\right)^{\gamma/(\gamma+1)}}-1\right]\label{eq:EMW}
\end{equation}
is achieved in the reversible case \citep{2020-finite-size}. And
$\eta_{\mathrm{C}}\equiv1-T_{c}^{[i]}/T_{h}^{[i]}$ is the Carnot
efficiency determined by the initial temperatures of the reservoirs.

Expressing $W_{d}$ in terms of $\eta$ according to Eq. (\ref{eq:eta-tau}),
the constraint on dissipative work ($W_{d}\geq W_{d}^{(\mathrm{min})}$)
becomes,

\begin{equation}
W_{d}=\frac{W_{\mathrm{max}}\left(\eta_{\mathrm{MW}}-\eta\right)}{\eta_{\mathrm{MW}}\left[1-\eta/(1+\gamma)\right]}\geq\frac{T_{b}\Sigma_{\mathrm{min}}}{\tau}.\label{eq:20}
\end{equation}
Eliminating the duration $\tau$ in this inequality with the average
power $P\equiv W(\tau)/\tau$ of the whole process, we find the trade-off
relation between power and efficiency \citep{S-M}

\begin{equation}
\tilde{P}\leq\frac{4\lambda\tilde{\eta}\left(1-\tilde{\eta}\right)}{\left(\lambda\tilde{\eta}+1-\tilde{\eta}\right)^{2}}.\label{eq:trade-off}
\end{equation}
Here, $\lambda\equiv1-\eta_{\mathrm{MW}}/(1+\gamma)$, $\tilde{P}\equiv P/P_{\mathrm{max}}$,
$\tilde{\eta}\equiv\eta/\eta_{\mathrm{MW}}$, and $P_{\mathrm{max}}\equiv W_{\mathrm{max}}^{2}/(4T_{b}\Sigma_{\mathrm{min}})$
is the maximum average power. 
\begin{figure}
\includegraphics[width=8.5cm]{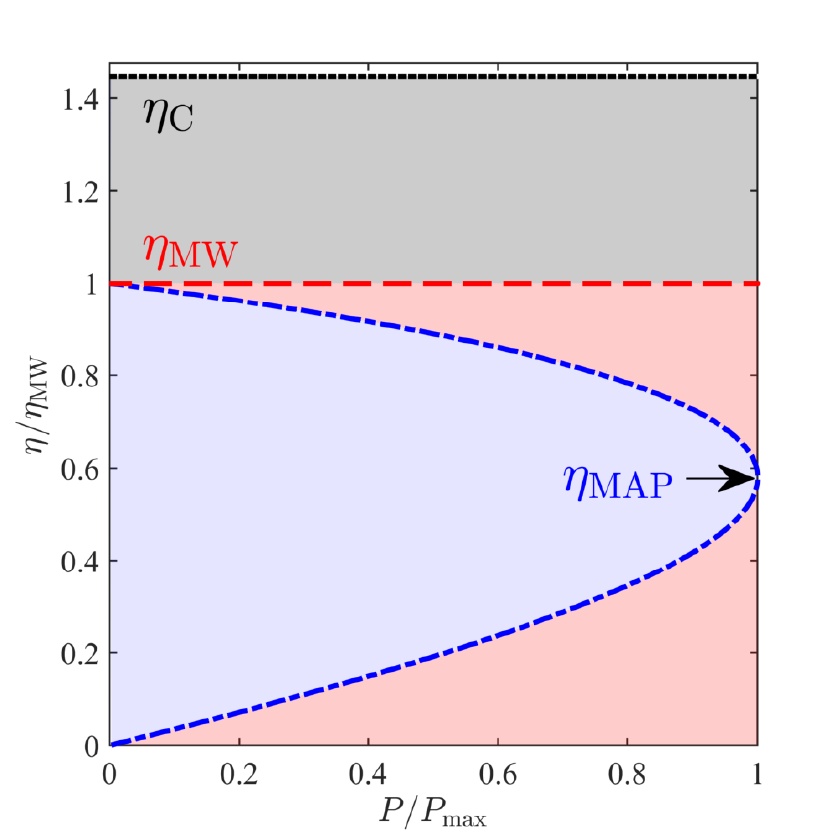}\caption{\label{fig:Trade-off-between-average}\textquotedbl Phase diagram\textquotedbl{}
$\tilde{P}-\tilde{\eta}$ of the heat engine performance between finite
reservoirs. The blue dash-dotted curve and the (light blue) area therein
represent the trade-off between $\tilde{P}=P/P_{\mathrm{max}}$ and
$\tilde{\eta}=\eta/\eta_{\mathrm{MW}}$ in Eq. (\ref{eq:trade-off}).
$P_{\mathrm{max}}$ is the maximum average power. The efficiency at
maximum work $\eta_{\mathrm{MW}}$ in Eq. (\ref{eq:EMW}) is plotted
with the red dashed line, while the corresponding Carnot efficiency
$\mathrm{\eta}_{\mathrm{C}}=0.8$ is plotted with the black dotted
line. In this example, we use $\gamma=C_{c}/C_{h}=1$.}
\end{figure}
As the main result of this paper, the above relation specifics the
complete optimization regime for the heat engines operating between
finite-sized reservoirs. The equal sign of Eq. (\ref{eq:trade-off})
is achieved with the minimum entropy generation $\sigma_{\mathrm{min}}$,
which determines the optimal performance of the heat engine, namely,
the maximum power for a given efficiency. The optimal operation of
the heat engine will be discussed later. We emphasize that such a
trade-off constrains the performance of all the heat engines operating
in the linear response regime, because $\sigma_{\mathrm{min}}$ is
the overall lower bound for irreversibility as we remarked below Eq.
(\ref{eq:sigma min}).

In the symmetric case with $\gamma=1$ (See SM \citep{S-M} for the
asymmetric cases with $\gamma=0.01,100$), the power-efficiency trade-off
is illustrated in Fig. \ref{fig:Trade-off-between-average} with the
blue dash-dotted curve and the (light blue) area therein. The efficiency
corresponding to the maximum power ($\tilde{P}=1$) is denoted as
$\eta_{\mathrm{MAP}}$ in this figure, and will be detailed discussed
in the following. $\eta_{\mathrm{C}}=0.8$ is used in this plot. Due
to the finiteness of the heat reservoirs, the (gray) area between
efficiency at maximum work $\eta_{\mathrm{MW}}$ (red dashed line)
and Carnot efficiency $\eta_{\mathrm{C}}$ (black dotted line) becomes
a forbidden regime in the \textquotedbl phase diagram\textquotedbl{}
of the heat engine performance. Particularly, in the limit of $\gamma\rightarrow\infty$
with infinite cold reservoir, the trade-off in Eq. (\ref{eq:trade-off})
reduces to a concise form $\tilde{P}\leq4\tilde{\eta}\left(1-\tilde{\eta}\right).$

With the obtained power-efficiency trade-off, it is straightforward
to find the efficiency at an arbitrary given power $\tilde{P}$ being
bounded in the region of $\tilde{\eta}_{-}\leq\tilde{\eta}\leq\tilde{\eta}_{+}$,
where $\tilde{\eta}_{\pm}$ are defined as \citep{S-M}
\begin{equation}
\tilde{\eta}_{\pm}\equiv1-\frac{\lambda\tilde{P}}{\left(1\pm\sqrt{1-\tilde{P}}\right)^{2}+\lambda\tilde{P}}.\label{eq:eta+-}
\end{equation}
The upper bound $\tilde{\eta}_{+}$, serving as the maximum efficiency
for an arbitrary average power, returns to its counterpart in the
infinite-reservoir case by replacing $\eta_{\mathrm{MW}}$ with $\eta_{\mathrm{C}}$
\citep{tradeoffholubec,long2016efficiency,Constraintrelationyhma}.
Obviously, $\tilde{\eta}_{+}$ approaches $1$ in the quasi-static
regime of $\tilde{P}\rightarrow0$, namely, $\eta\rightarrow\eta_{\mathrm{MW}}$,
as shown in Fig. \ref{fig:Trade-off-between-average}.

\emph{Efficiency at maximum average power.} -- When the heat engine
achieve its maximum average power ($\tilde{P}=1$), the upper and
lower bound in Eq. (\ref{eq:eta+-}) converge to the efficiency at
maximum average power (EMAP)

\begin{equation}
\eta_{\mathrm{MAP}}=\frac{\eta_{\mathrm{MW}}}{2-\eta_{\mathrm{MW}}/\left(1+\gamma\right)}.\label{eq:EMAP}
\end{equation}
We note that this general EMAP recovers $\eta_{\mathrm{MW}}/2$ ($\gamma\rightarrow\infty$)
which was previously obtained in the special case with infinite large
cold reservoir \citep{izumida2014work}. Since $\gamma\in[0,\infty]$,
$\eta_{\mathrm{MAP}}$ satisfies

\begin{equation}
\eta_{\mathrm{L}}\equiv\frac{\eta_{\mathrm{MW}}}{2}\leq\eta_{\mathrm{MAP}}\leq\frac{\eta_{\mathrm{MW}}}{2-\eta_{\mathrm{MW}}}\equiv\eta_{\mathrm{U}},\label{eq:boundEMAP}
\end{equation}
where the upper bound $\eta_{\mathrm{U}}$ is reached in the limit
$\gamma\rightarrow0$ ($C_{c}\ll C_{h}$) with infinite large hot
reservoir.

\begin{figure}
\begin{raggedright}
(a)
\par\end{raggedright}
\begin{raggedright}
\includegraphics[width=8.5cm]{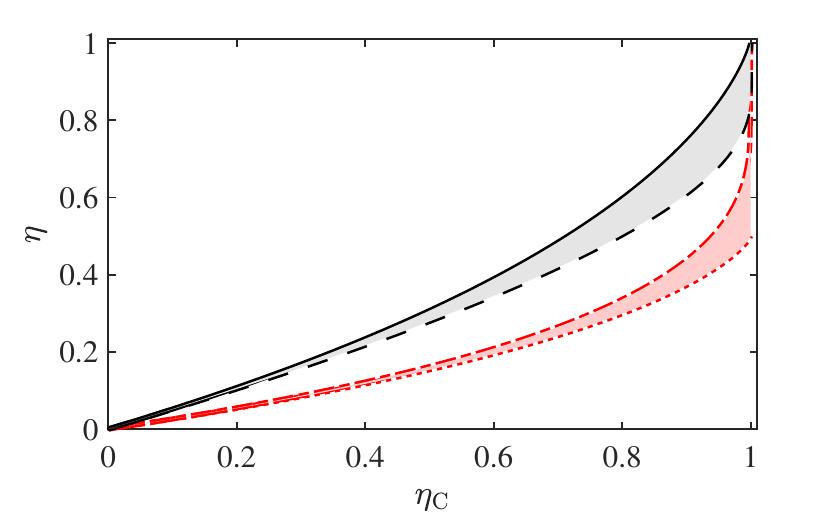}
\par\end{raggedright}
\begin{raggedright}
(b)
\par\end{raggedright}
\begin{raggedright}
\includegraphics[width=8.5cm]{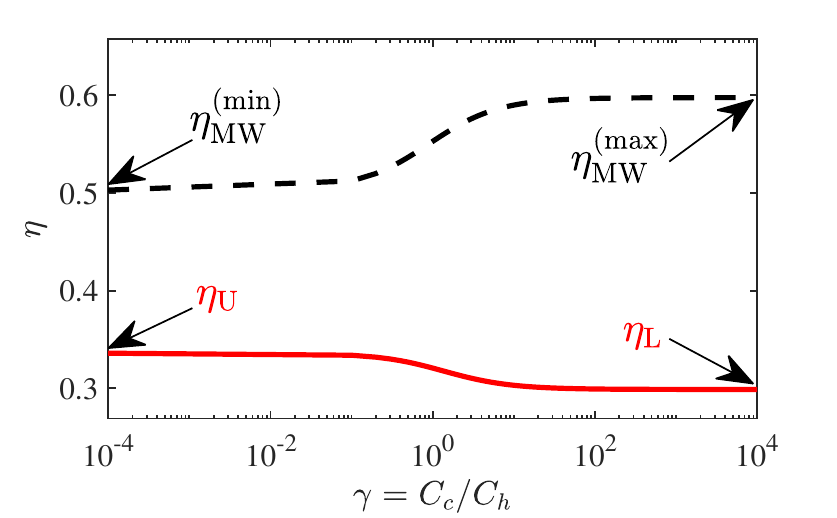}
\par\end{raggedright}
\caption{\label{fig:Dependence-of-and}Dependence of and $\eta_{\mathrm{MAP}}$
and $\eta_{\mathrm{MW}}$ on $\eta_{\mathrm{C}}$ and $\gamma$. (a)
$\eta_{\mathrm{MAP}}$ and $\eta_{\mathrm{MW}}$ as the function of
$\eta_{\mathrm{C}}$. The upper (lower) bound $\eta_{\mathrm{U}}$
($\eta_{\mathrm{L}}$) of $\eta_{\mathrm{MAP}}$ in Eq. (\ref{eq:boundEMAP})
is plotted as the red dash-dotted (dotted) curve, the (light red)
area between the dash-dotted curve and dotted curve is the available
range of $\eta_{\mathrm{MAP}}$. The upper (lower) bound of $\eta_{\mathrm{MW}}$
is represented by the black solid (dashed) curve, and the (gray) area
between the solid curve and dashed curve is the achievable range of
$\eta_{\mathrm{MW}}$. (b) $\eta_{\mathrm{MAP}}$ and $\eta_{\mathrm{MW}}$
as the function of $\gamma$. The red solid curve and black dashed
curve represent $\eta_{\mathrm{MAP}}$ and $\eta_{\mathrm{MW}}$,
respectively. In this example, $\eta_{\mathrm{C}}=0.8$.}
\end{figure}

Figure \ref{fig:Dependence-of-and}(a) shows the dependence of $\eta_{\mathrm{MAP}}$
on $\eta_{\mathrm{C}}$, where the (light red) area between $\eta_{\mathrm{U}}$
(red dash-dotted curve) and $\eta_{\mathrm{L}}$ (red dotted curve)
is the available range of $\eta_{\mathrm{MAP}}$. In comparison, the
achievable range of $\eta_{\mathrm{MW}}$ is represented with the
(gray) area between the black solid curve and the black dashed curve.
As demonstrated in this figure, in the small-$\eta_{\mathrm{C}}$
regime, there exist $\gamma-$independent scalings for $\eta_{\mathrm{MAP}}$
and $\eta_{\mathrm{MW}}$. Such universalities can be explicitly obtained
by expanding $\eta_{\mathrm{MAP}}$ and $\eta_{\mathrm{MW}}$ with
respect to $\eta_{\mathrm{C}}$:

\begin{equation}
\eta_{\mathrm{MW}}=\frac{1}{2}\eta_{\mathrm{C}}+\frac{1}{6}\left(1-\frac{1/2}{\gamma+1}\right)\eta_{\mathrm{C}}^{2}+\mathcal{O}\left(\eta_{\mathrm{C}}^{3}\right),\label{eq:30}
\end{equation}

\begin{equation}
\eta_{\mathrm{MAP}}=\frac{1}{4}\eta_{\mathrm{C}}+\frac{1}{12}\left(1+\frac{1/4}{\gamma+1}\right)\mathrm{\eta}_{\mathrm{C}}^{2}+\mathcal{O}\left(\eta_{\mathrm{\mathrm{C}}}^{3}\right).\label{eq:31}
\end{equation}
Obviously, the first-order coefficients of both $\eta_{\mathrm{MW}}$
and $\eta_{\mathrm{MAP}}$ are independent of the heat capacity ratio
$\gamma$, as we inferred from Fig. \ref{fig:Dependence-of-and}(a).
Up to the first order of $\eta_{\mathrm{C}}$, the universality of
$\eta_{\mathrm{MAP}}$ scales as $\eta_{\mathrm{MAP}}\sim\eta_{\mathrm{C}}/4$.
Meanwhile, the universality of $\eta_{\mathrm{MW}}$ follows as $\eta_{\mathrm{MW}}\sim\eta_{\mathrm{C}}/2$,
which has also been revealed in previous studies \citep{johal2016near,2020-finite-size}.
Nevertheless , the coefficients corresponding to the second order
of $\eta_{\mathrm{C}}$ are $\gamma-$dependent for $\eta_{\mathrm{MAP}}$
and $\eta_{\mathrm{MW}}$. The signs of the terms containing $\gamma$
in $\eta_{\mathrm{MAP}}$ and $\eta_{\mathrm{MW}}$ are opposite,
which indicates that the monotonicity of $\eta_{\mathrm{MAP}}$ and
$\eta_{\mathrm{MW}}$ with respect to $\gamma$ is opposite. This
fact is clearly illustrated in Fig. $\text{\ref{fig:Dependence-of-and}}$(b),
where $\eta_{\mathrm{MAP}}$ (red solid curve) is a monotonically
decreasing function of $\gamma$, while $\eta_{\mathrm{MW}}$ (black
dashed curve) increases with $\gamma$ monotonically \citep{johal2016near}.
In this figure, the maximum $\eta_{\mathrm{MW}}$ ($\eta_{\mathrm{MW}}^{(\mathrm{max})}$)
and minimum $\eta_{\mathrm{MW}}$ ($\eta_{\mathrm{MW}}^{(\mathrm{min})}$)
are reached in the limit $\gamma\rightarrow\infty$ and $\gamma\rightarrow0$,
respectively \citep{2020-finite-size}, and $\eta_{\mathrm{C}}=0.8$
is fixed. As the result of the opposite monotonicity, there exists
a competitive relation between $\eta_{\mathrm{MAP}}$ and $\eta_{\mathrm{MW}}$.
Namely, $\eta_{\mathrm{MAP}}$ achieves its maximum even when $\eta_{\mathrm{MW}}$
is minimum in the limit $\gamma\rightarrow0$, and vice versa. 

\textit{Optimal operation protocol of the heat engine}. -- As a process
function, the path dependence of entropy production $\sigma$ in the
parameter space makes it relies on the control protocol applied to
the working substance \citep{yhmaoptimalcontrol,2020IEGyhma}. Therefore,
the efficiency and power of the heat engine are inseparable from the
specific operation protocol of the cycle. To achieve the boundary
of the trade-off (\ref{eq:trade-off}) or the EMAP (\ref{eq:EMAP}),
we demonstrate the optimal operation of the heat engine associated
with the minimal entropy production $\sigma_{\mathrm{min}}$ with
a specific example. For a finite-time Carnot heat engine whose working
substance is the ideal gas with volume $V$ (control parameter), the
minimal entropy production condition, i.e., $\dot{Q}_{h}=\sqrt{L_{22}\Sigma_{\mathrm{min}}}/\tau$,
allows us to find the optimal control protocol for $V(t)$ from the
energy conservation relation of the gas \citep{S-M}.

\begin{figure}
\includegraphics[width=8.5cm]{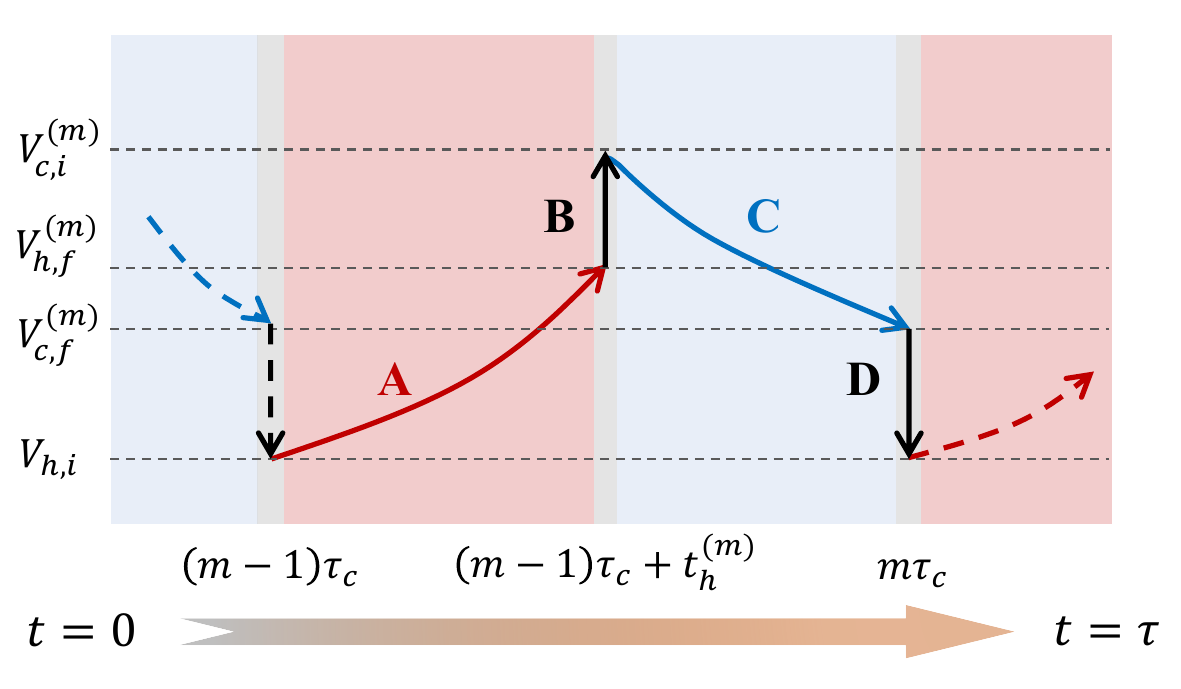}\caption{\label{fig:The-diagram-of-operation}The diagram of the optimal operation
protocol of the $m$-th cycle with the control parameter $V$ (gas
volume). In the isothermal expansion (expression) process $\mathrm{A}$
($\mathrm{C}$) of duration $t_{h}$ ($t_{c}$), $V$ changes exponentially
with time; while in adiabatic processes ($\mathrm{B}$ and $\mathrm{D}$),
$V$ is quenched with the adiabatic equation of idea gas being satisfied.}
\end{figure}

The optimal operation protocol of the heat engine is shown in Fig.
\ref{fig:The-diagram-of-operation}, where $\mathrm{A}$ ($\mathrm{C}$)
represents the finite-time isothermal expansion (compression) process
with duration $t_{h}^{(m)}$($t_{c}^{(m)}$) in the $m$-th ($m=1,2,3...M$)
cycle. During the isothermal expansion (compression), the gas volume
changes exponentially with time as $V_{h}^{(m)}(\tilde{t})=V_{h,i}^{(m)}\exp(\varGamma_{h}^{(m)}\tilde{t})$
($V_{c}^{(m)}(t')=V_{c,i}^{(m)}\exp(-\varGamma_{c}^{(m)}t')$) with
$\tilde{t}\equiv t-(m-1)\tau_{c}$ ($t'\equiv t-(m-1)\tau_{c}-t_{h}^{(m)}$).
Here, the initial volume of the gas in the isothermal expansion process
$V_{h,i}^{(m)}=V_{h,i}$ is fixed in each cycle, while the initial
volume of other three processes ($V_{h,f}^{(m)}$, $V_{c,i}^{(m)}$,
and $V_{c,f}^{(m)}$) are determined by the full operation protocol.
$\varGamma_{h(c)}^{(m)}$ represents the isothermal expansion (compression)
rate of the $m$-th cycle \citep{S-M}. The adiabatic equation of
ideal gas is satisfied in the adiabatic processes $\mathrm{B}$ and
$\mathrm{D}$, the duration of which is ignored in comparison with
that of the isothermal processes \citep{EspositoPRL2010,ZCTuCPB,2020IEGyhma}.
It is worth mentioning that a recent study \citep{CAengine-Quan}
obtained similar optimal operation to realize the efficiency at maximum
power of a Brownian heat engine between constant temperature reservoirs.
This reminds us that such an optimal operation scheme may be universal
for some types of finite-time heat engines. 

\emph{Conclusion and discussion--}In summary, we successfully obtained
a general power-efficiency trade-off for heat engines operating between
two finite-sized reservoirs within the linear response regime. With
such a trade-off, we showed the achievable range of efficiency for
a given average power, and the universal efficiency at maximum average
power. To achieve the optimal performance of the heat engine, corresponding
to the boundary of the power-efficiency trade-off, the optimal operation
protocol of an ideal gas heat engine is demonstrated. The predicted
results can be tested on some state-of-art platforms \citep{BrownianHENatPhys2015,2020IEGyhma}.
Moreover, by replacing $\eta_{\mathrm{MW}}$ with $\eta_{\mathrm{C}}$,
some typical constraints in finite case become their corresponding
counterparts in infinite case. These thermodynamic constraints specific
the full operation regime of the heat engines in real-world circumstances.
Basically, this study paves the way for the joint optimization of
thermodynamic cycle by adjusting the ratio of the heat capacities
of the reservoirs and controlling the operation of the cycle, and
may shed new light on investigating the irreversibility of non-equilibrium
thermodynamic processes off thermodynamic limit. 

The temperature-dependent feature of the reservoir's heat capacity
\citep{2020-finite-size}, the quantumness of the reservoir \citep{xu2014noncanonical,NanoscaleHEPRL2014,quantum2nd-2021ma},
the deviation of entropy production from $1/\tau$-scaling beyond
the slow-driving regime \citep{Constraintrelationyhma,yhmaoptimalcontrol,2020IEGyhma,Fast-drivingMa},
and the fluctuations in heat engine performance \citep{2014Universal-flu,2020Efficiency-flu,eff-fluctu-2021fei}
are expected to be taken into future considerations.

\textit{Acknowledgment}.-We thank G. H. Dong and Y. Chen for the helpful
suggestions. We are grateful to the anonymous referees for enlightening
comments. This work is supported by the National Science Foundation
of China (NSFC) (Grants No. 11534002, No. 11875049, No. U1730449,
No. U1530401, and No. U1930403), the National Basic Research Program
of China (Grants No. 2016YFA0301201), and the China Postdoctoral Science
Foundation (Grant No. BX2021030).

\bibliographystyle{apsrev}
\addcontentsline{toc}{section}{\refname}\bibliography{TOFS}

\end{document}


\title{\textit{\Large{}Supplementary Materials for}{\Large{} \textquotedbl Optimizing
Thermodynamic cycles with Finite-Sized Reservoirs\textquotedbl}}
\author{Hong Yuan}
\address{Graduate School of China Academy of Engineering Physics, No. 10 Xibeiwang
East Road, Haidian District, Beijing, 100193, China}
\author{Yu-Han Ma}
\email{yhma@gscaep.ac.cn}

\address{Graduate School of China Academy of Engineering Physics, No. 10 Xibeiwang
East Road, Haidian District, Beijing, 100193, China}
\author{C. P. Sun}
\email{suncp@gscaep.ac.cn}

\address{Graduate School of China Academy of Engineering Physics, No. 10 Xibeiwang
East Road, Haidian District, Beijing, 100193, China}
\address{Beijing Computational Science Research Center, Beijing 100193, China}

\maketitle
This document is devoted to providing the detailed derivations and
the supporting discussions to the main content of the Letter. In Sec.
\ref{sec:The-reversible-regime}, we discuss the final uniform temperature
$\tilde{T}$, the maximum work $W_{\mathrm{max}}$, and the corresponding
efficiency $\eta_{\mathrm{MW}}$ in the reversible regime under the
quasi-static limit. The lower bound of irreversible entropy production,
illustrated in {[}Eq. (5){]} of the main text, is derived in Sec.
\ref{sec:Lower-bound-of}. In Sec. \ref{sec:Dissipation-work-and},
we show the proofs of the lower bound of dissipative work $W_{d}\geq T_{b}\Sigma_{\mathrm{min}}/\tau$,
and the trade-off between power and efficiency presented in {[}Eq.
(10){]} of the main text. The derivation of the bounds for efficiency
at arbitrary given power $\tilde{\eta}_{\pm}$ {[}Eq. (11) of the
main text{]} is given in Sec. \ref{sec:Bounds-of-efficiency}. In
Sec. \ref{sec:Optimal-operation-protocol}, taking idea gas heat engine
as an example, we demonstrate the optimal operation protocol of the
engine associated with the minimal entropy production to achieve the
boundary of the power-efficiency trade-off.

\section{\label{sec:The-reversible-regime}The reversible regime}

The extractable work from the reservoirs in the reversible regime,
$W_{\mathrm{max}}\equiv\underset{\sigma\rightarrow0}{\lim}W(\tau)$,
is the upper bound of the work output until the heat engines stop
operating. Such bound is achieved with no irreversible entropy production
in the whole process \citep{ondrechen1981maximum,izumida2014work,2020-finite-size}.
Specifically, in an infinitesimal process, The heat engine absorbs
$dQ_{h}=-C_{h}dT_{h}$ from the hot reservoir, and then transforms
it into the infinitesimal work the $dW_{\mathrm{max}}$ with the corresponding
time-dependent Carnot efficiency $\eta_{\mathrm{C}}(t)\equiv1-T_{c}/T_{h}$
\citep{izumida2014work}. In this sense, the power of the engine reads
\begin{equation}
dW_{\mathrm{max}}=-C_{h}dT_{h}(1-\frac{T_{c}}{T_{h}}).\label{eq:(1)}
\end{equation}
According to the conservation of energy $dW_{\mathrm{max}}=dQ_{h}-dQ_{c}$,
we can also write 
\begin{equation}
dW_{\mathrm{max}}=-C_{h}dT_{h}-C_{c}dT_{c},\label{eq:(2)}
\end{equation}
where $dQ_{c}=C_{c}dT_{c}$ is the heat released from the heat engine
to the cold reservoir. Comparing Eq. ($\text{\ref{eq:(1)}}$) and
Eq. ($\text{\ref{eq:(2)}}$), one has
\begin{equation}
-C_{h}\frac{dT_{h}}{T_{h}}=C_{c}\frac{dT_{c}}{T_{c}}.\label{eq:(3)}
\end{equation}
In the case with constant heat capacity $C_{h}(C_{c})$, by integrating
both sides of Eq. (\ref{eq:(3)}), the final uniform temperature of
the total system, $T_{b}\equiv T_{c}^{[f]}=T_{h}^{[f]}$, is obtained
as 
\begin{equation}
T_{b}=\left[T_{h}^{[i]}\right]^{\frac{1}{\gamma+1}}\left[T_{c}^{[i]}\right]^{\frac{\gamma}{\gamma+1}},\label{eq:(6)}
\end{equation}
where $\gamma\equiv C_{c}/C_{h}$ is defined as the heat capacity
ratio. As a consequence, integrating Eq. ($\text{\ref{eq:(2)}}$)
, we find the maximum work output 
\begin{equation}
W_{\mathrm{max}}=C_{h}\left(T_{h}^{[i]}-T_{b}\right)-C_{c}\left(T_{b}-T_{c}^{[i]}\right).\label{eq:Wmax}
\end{equation}
The first term of the right hand of Eq. (\ref{eq:Wmax}) is the reversible
heat absorbed, $Q_{h}\left(\sigma\rightarrow0\right)=C_{h}\left[T_{h}^{[i]}-T_{b}\right]$,
achieved in the quasi-static limit. In this case, the corresponding
efficiency, namely, the efficiency at maximum work output ($\mathrm{EMW}$)
$\eta_{\mathrm{MW}}\equiv W_{\mathrm{max}}/Q_{h}\left(\sigma\rightarrow0\right)$,
is directly obtained as \citep{2020-finite-size}
\begin{equation}
\eta_{\mathrm{MW}}=1-\gamma\left[\frac{\eta_{\mathrm{C}}}{1-\left(1-\eta_{\mathrm{C}}\right)^{\frac{\gamma}{\gamma+1}}}-1\right],\label{eq:EMW}
\end{equation}
where $\eta_{\mathrm{C}}\equiv\eta_{\mathrm{C}}(0)=1-T_{c}^{[i]}/T_{h}^{[i]}$
is the Carnot efficiency determined by the initial temperature of
the two reservoirs. As we mentioned before, this result is applicable
to the heat reservoirs with constant heat capacity. For the reservoirs
with temperature-dependent heat capacity, the generalization of Eq.
(\ref{eq:EMW}) is studied our previous work \citep{2020-finite-size}.

\section{\label{sec:Lower-bound-of}Lower bound of irreversible entropy production}

It follows from {[}Eq. (3){]} and {[}Eq. (4){]} of the main text that
the entropy production $\sigma(\tau)=\int_{0}^{\tau}(\sqrt{\dot{\sigma}})^{2}dt$
explicitly follows as

\begin{align}
\sigma(\tau) & \geq\frac{\left(\int_{0}^{\tau}C_{h}\dot{T}_{h}/\sqrt{L_{22}}dt\right)^{2}}{\tau},\label{eq:}\\
 & =\frac{\left(\int_{T_{h}^{[i]}}^{\tilde{T}}C_{h}dT_{h}/\sqrt{L_{22}}\right)^{2}}{\tau}\\
 & \equiv\frac{\Sigma}{\tau}\label{eq:sigma min-1}
\end{align}
where $\Sigma=\Sigma(\tilde{T})$, and $T_{h}(\tau)\equiv\tilde{T}=\tilde{T}(\sigma)$
is given by {[}Eq. (2){]} of the main text. In the linear irreversible
regime, keeping the first order of $\sigma$ in $\tilde{T}(\sigma)$
, {[}Eq. (2){]} of the main text is approximated as

\begin{align}
\tilde{T} & =T_{b}\exp\left[\frac{\sigma}{C_{h}+C_{c}}\right]\\
 & \approx T_{b}\left[1+\frac{\sigma}{C_{h}+C_{c}}\right].
\end{align}
Then, $\Sigma$ is approximated as

\begin{align}
\Sigma & =\Sigma(\tilde{T}=T_{b})+\left.\frac{\partial\Sigma}{\partial\tilde{T}}\right|_{\tilde{T}=T_{b}}\left(\tilde{T}-T_{b}\right)\\
 & =\left(\int_{T_{h}^{[i]}}^{T_{b}}C_{h}dT_{h}/\sqrt{L_{22}}\right)^{2}+2\left(\int_{T_{h}^{[i]}}^{T_{b}}C_{h}dT_{h}/\sqrt{L_{22}}\right)\frac{T_{b}\sigma}{\sqrt{L_{22}}\left(1+\gamma\right)},
\end{align}
up to the first order of $\sigma$. Substituting the above result
into Eq. (\ref{eq:sigma min-1}), we obtain

\begin{equation}
\sigma(\tau)\geq\frac{\Sigma_{\mathrm{min}}}{\tau}\left[1-\frac{2T_{b}\sqrt{\Sigma_{\mathrm{min}}/L_{22}}}{\left(1+\gamma\right)\tau}\right]^{-1},
\end{equation}
where $\Sigma_{\mathrm{min}}\equiv\left(\int_{T_{h}^{[i]}}^{T_{b}}C_{h}dT_{h}/\sqrt{L_{22}}\right)^{2}$
is a $\tau$-independent dissipation coefficient. By expanding the
right hand of this inequality with respect to $\tau^{-1}$, we have

\begin{equation}
\sigma(\tau)\geq\frac{\Sigma_{\mathrm{min}}}{\tau}\left[1+\frac{2T_{b}\sqrt{\Sigma_{\mathrm{min}}/L_{22}}}{\left(1+\gamma\right)\tau}\right]+\mathcal{O}\left(\tau^{-3}\right).
\end{equation}
In the long-time regime, up to the first order of $\tau^{-1}$, we
obtain {[}Eq. (5){]} of the main text, namely, $\sigma(\tau)\geq\Sigma_{\mathrm{min}}/\tau$.
To be consistent with this result, all the high-order terms of $\tau^{-1}$
will be ignored in the following discussion.

It is worth mentioning that, even without the tight-coupling condition,
{[}Eq.(5){]} of the main text serves as the overall lower bound for
entropy production. The proof is as follows: based on the linear irreversible
thermodynamics, the entropy production rate $\dot{\sigma}$ reads
\citep{wang2014optimization}

\begin{equation}
\dot{\sigma}=\frac{1}{L_{22}q^{2}}J_{2}^{2}+\frac{1-q^{2}}{q^{2}}L_{22}X_{2}^{2}-2\frac{1-q^{2}}{q^{2}}J_{2}X_{2}.
\end{equation}
Here $J_{2}\equiv\dot{Q_{h}}$ is the heat flow from the hot reservoir,
$X_{2}\equiv1/T_{c}-1/T_{h}$ represents the conjugate thermodynamic
force, and the coefficient of the coupling strength $q\equiv L_{21}/\sqrt{L_{11}L_{22}}$
satisfies $\lvert q\rvert\le1$. The derivative of $\sigma$ with
respect to $q^{2}$ is straightforward obtained as

\begin{align}
\frac{\partial\sigma}{\partial\left(q^{2}\right)} & =\int_{0}^{\tau}\frac{\partial\dot{\sigma}}{\partial\left(q^{2}\right)}dt\\
 & =-\int_{0}^{\tau}\frac{1}{q^{4}}\left(\frac{J_{2}}{\sqrt{L_{22}}}-\sqrt{L_{22}}X_{2}\right)^{2}dt\leq0,
\end{align}
which is always negative. This indicates that the entropy production
decreases monotonically with the increase of $\lvert q\rvert$. Therefore,
the minimum entropy production obtained in {[}Eq.(5){]} of the main
text serves as the overall lower bound for entropy production with
arbitrary $q$.

\section{\label{sec:Dissipation-work-and}Dissipative work and power-efficiency
trade-off}

\subsection{The dissipative work}

The work output $W(\tau)=Q_{h}(\tau)-Q_{c}(\tau)$ is explicitly written
as

\begin{equation}
W(\tau)=C_{h}\left(T_{h}^{[i]}-\tilde{T}\right)-C_{c}\left(\tilde{T}-T_{c}^{[i]}\right),\label{eq:Wtau}
\end{equation}
which is monotonically decreasing as $\tilde{T}$ increases. It follows
from Eqs. (\ref{eq:Wmax}) and (\ref{eq:Wtau}) that the dissipative
work $W_{d}(\tau)\equiv W_{\mathrm{max}}-W(\tau)$ reads

\begin{equation}
W_{d}(\tau)=\left(C_{c}+C_{h}\right)\left(\tilde{T}-T_{b}\right).\label{eq:Wd-1}
\end{equation}
Combining {[}Eq. (2){]} and {[}Eq. (5){]} of the main text, one has

\begin{equation}
\tilde{T}\geq T_{b}\exp\left[\frac{\Sigma_{\mathrm{min}}}{\left(C_{c}+C_{h}\right)\tau}\right],\label{eq:Tfinal-1}
\end{equation}
and then Eq. (\ref{eq:Wd-1}) is written in terms of $\tau$ as,

\begin{align}
W_{d}(\tau) & \geq\left(C_{h}+C_{c}\right)T_{b}\left[e^{\frac{\Sigma_{\mathrm{min}}}{\left(C_{c}+C_{h}\right)\tau}}-1\right]\\
 & =\frac{T_{b}\Sigma_{\mathrm{min}}}{\tau}+\mathcal{O}\left(\tau^{-2}\right).
\end{align}
Up to the first order of $\tau^{-1}$, the inequality for the dissipative
work in main text is obtained as $W_{d}(\tau)\geq T_{b}\Sigma_{\mathrm{min}}/\tau$.

\subsection{Power-efficiency trade-off}

\begin{figure}
\centering{}\subfloat[]{\begin{centering}
\includegraphics[width=8.5cm]{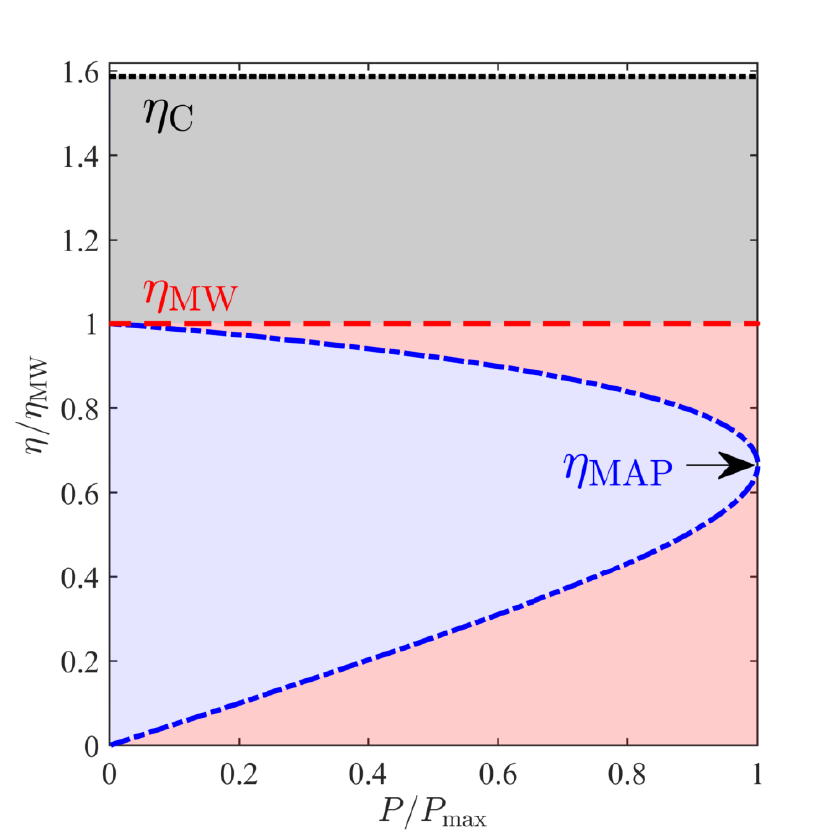}
\par\end{centering}
}\subfloat[]{\begin{centering}
\includegraphics[width=8.5cm]{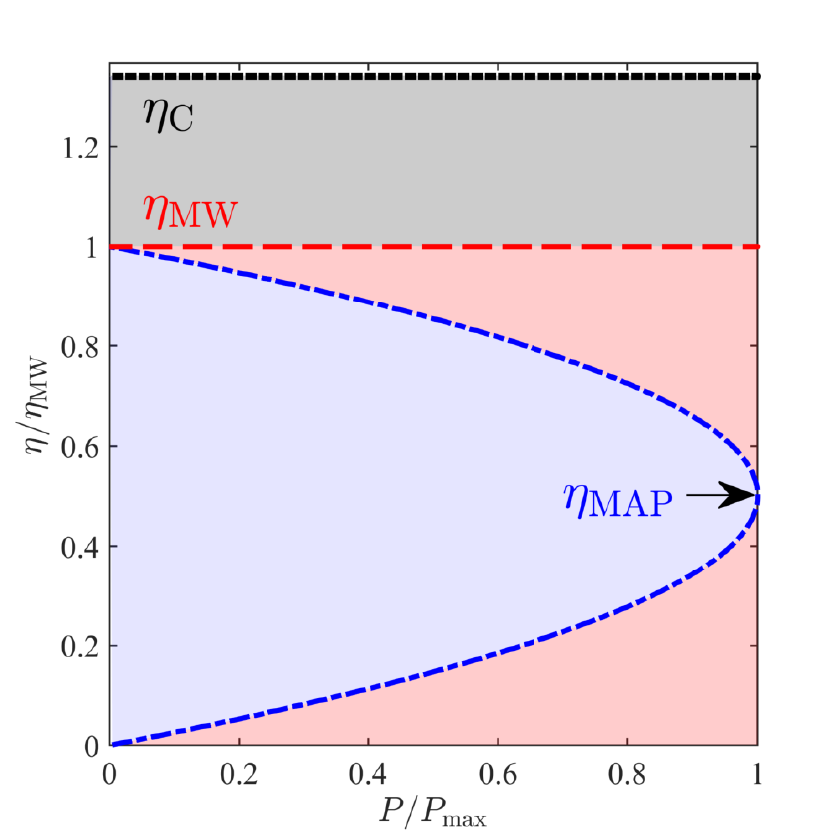}
\par\end{centering}
}\caption{\label{fig:"Phase-diagram"-1}\textquotedbl Phase diagram\textquotedbl{}
$\tilde{P}-\tilde{\eta}$ of the heat engine performance between finite
reservoirs with different heat capacity ratio $\gamma=0.01$ {[}(a){]}
and $\gamma=100$ {[}(b){]}. The blue dash-dotted curve and the (light
blue) area therein represent the trade-off between $\tilde{P}=P/P_{\mathrm{max}}$
and $\tilde{\eta}=\eta/\eta_{\mathrm{MW}}$ in Eq. (\ref{eq:constraint-relation}),
where $P_{\mathrm{max}}$ is the maximum average power. The efficiency
at maximum work $\eta_{\mathrm{MW}}$ in Eq. (\ref{eq:EMW}) is plotted
with the red dashed line, while the corresponding Carnot efficiency
$\mathrm{\eta}_{\mathrm{C}}=0.8$ is plotted with the black dotted
line.}
\end{figure}

As shown in {[}Eq. (9){]} of the main text that

\begin{equation}
\frac{\eta_{\mathrm{MW}}-\eta}{\eta_{\mathrm{MW}}\left[1-\eta/(1+\gamma)\right]}\geq\frac{T_{b}\Sigma_{\mathrm{min}}}{W_{\mathrm{max}}\tau},\label{eq:20-1}
\end{equation}
which becomes, by multiplying both sides by $W(\tau)$,

\begin{align}
\frac{W(\tau)\left(\eta_{\mathrm{MW}}-\eta\right)}{\eta_{\mathrm{MW}}\left[1-\eta/(1+\gamma)\right]} & \geq\frac{T_{b}\Sigma_{\mathrm{min}}W(\tau)}{W_{\mathrm{max}}\tau}\label{eq:-1}\\
 & =\frac{T_{b}\Sigma_{\mathrm{min}}P}{W_{\mathrm{max}}}.\label{eq:inequal}
\end{align}
Here, $P\equiv W(\tau)/\tau$ is defined as the average power of the
whole process. Further, we rewrite $W(\tau)$ as

\begin{align}
W(\tau) & =W_{\mathrm{max}}-W_{d}\\
 & =W_{\mathrm{max}}-\frac{W_{\mathrm{max}}\left(\eta_{\mathrm{MW}}-\eta\right)}{\eta_{\mathrm{MW}}\left[1-\eta/(1+\gamma)\right]},
\end{align}
then Eq. (\ref{eq:inequal}) becomes

\begin{equation}
W_{\mathrm{max}}\left\{ 1-\frac{\eta_{\mathrm{MW}}-\eta}{\eta_{\mathrm{MW}}\left[1-\eta/(1+\gamma)\right]}\right\} \frac{\left(\eta_{\mathrm{MW}}-\eta\right)}{\eta_{\mathrm{MW}}\left[1-\eta/(1+\gamma)\right]}\geq\frac{T_{b}\Sigma_{\mathrm{min}}P}{W_{\mathrm{max}}}.
\end{equation}
By straightforward calculation, the above inequality is simplified
as

\begin{equation}
P\leq\frac{W_{\mathrm{max}}^{2}}{\eta_{\mathrm{MW}}^{2}T_{b}\Sigma_{\mathrm{min}}}\frac{\eta\left[1-\eta_{\mathrm{MW}}/(1+\gamma)\right]\left(\eta_{\mathrm{MW}}-\eta\right)}{\left[1-\eta/(1+\gamma)\right]^{2}}.\label{eq:constraint1}
\end{equation}
On the other hand, the average power $P=W(\tau)/\tau$ follows as

\begin{align}
P & =\frac{W_{\mathrm{max}}-W_{d}}{\tau}\\
 & \leq\frac{W_{\mathrm{max}}-T_{b}\Sigma_{\mathrm{min}}/\tau}{\tau},
\end{align}
where $W_{d}\geq T_{b}\Sigma_{\mathrm{min}}/\tau$ has been used.
The above relation can be further written in the quadratic form, namely,

\begin{equation}
P\leq-T_{b}\Sigma_{\mathrm{min}}\left(\frac{1}{\tau}-\frac{W_{\mathrm{max}}}{2T_{b}\Sigma_{\mathrm{min}}}\right)^{2}+\frac{W_{\mathrm{max}}^{2}}{4T_{b}\Sigma_{\mathrm{min}}},
\end{equation}
which clearly shows that the maximum average power and the corresponding
optimal operation time are

\begin{equation}
P=\frac{W_{\mathrm{max}}^{2}}{4T_{b}\Sigma_{\mathrm{min}}}\equiv P_{\mathrm{max}},\;\mathrm{and\;}\tau=\frac{2T_{b}\Sigma_{\mathrm{min}}}{W_{\mathrm{max}}}\equiv\tau^{*},
\end{equation}
respectively. Finally, with the definitions

\begin{equation}
\tilde{P}\equiv\frac{P}{P_{\mathrm{max}}},\tilde{\eta}\equiv\frac{\eta}{\eta_{\mathrm{MW}}},\lambda\equiv1-\frac{\eta_{\mathrm{MW}}}{1+\gamma},
\end{equation}
Eq. (\ref{eq:constraint1}) is re-expressed as a concise form

\begin{equation}
\tilde{P}\leq\frac{4\lambda\tilde{\eta}\left(1-\tilde{\eta}\right)}{\left(\lambda\tilde{\eta}+1-\tilde{\eta}\right)^{2}}.\label{eq:constraint-relation}
\end{equation}
This is the trade-off between power and efficiency as we presented
in {[}Eq. (10){]} of the main text. In {[}Fig. 2{]} of the main text,
we plot such trade-off in the symmetric case with $\gamma=C_{c}/C_{h}=1$.
As comparisons, the trade-off in the asymmetric cases with $\gamma=0.01,100$
are shown in Fig. \ref{fig:"Phase-diagram"-1}.

\section{\label{sec:Bounds-of-efficiency}Bounds of efficiency at arbitrary
given power}

Note that the numerator in the right side of the trade-off in Eq.
(\ref{eq:constraint-relation}) can be re-expressed as

\begin{equation}
4\lambda\tilde{\eta}\left(1-\tilde{\eta}\right)=\left(\lambda\tilde{\eta}+1-\tilde{\eta}\right)^{2}-\left[\lambda\tilde{\eta}-\left(1-\tilde{\eta}\right)\right]^{2}.
\end{equation}
Thus, the trade-off is rewritten as

\begin{equation}
\tilde{P}+\left[\frac{\lambda\tilde{\eta}-\left(1-\tilde{\eta}\right)}{\lambda\tilde{\eta}+1-\tilde{\eta}}\right]^{2}\leq1,
\end{equation}
such that the upper and lower bounds in $\tilde{\eta}_{-}\leq\tilde{\eta}\leq\tilde{\eta}_{+}$
are the solutions of the quadratic equation

\begin{equation}
\left[\frac{\lambda\tilde{\eta}-\left(1-\tilde{\eta}\right)}{\lambda\tilde{\eta}+1-\tilde{\eta}}\right]^{2}=1-\tilde{P}.
\end{equation}
By straightforward calculation, one has

\begin{equation}
\tilde{\eta}_{\pm}=\frac{1\pm\sqrt{1-\tilde{P}}}{1\pm\sqrt{1-\tilde{P}}+\lambda\left(1\mp\sqrt{1-\tilde{P}}\right)}=1-\frac{\lambda\left(1\mp\sqrt{1-\tilde{P}}\right)}{1\pm\sqrt{1-\tilde{P}}+\lambda\left(1\mp\sqrt{1-\tilde{P}}\right)}.\label{eq:etaul}
\end{equation}
With the help of the relation

\begin{equation}
\tilde{P}=1-\left(\sqrt{1-\tilde{P}}\right)^{2}=\left(1\mp\sqrt{1-\tilde{P}}\right)\left(1\pm\sqrt{1-\tilde{P}}\right),
\end{equation}
Eq. (\ref{eq:etaul}) is simplified as

\begin{equation}
\tilde{\eta}_{\pm}=1-\frac{\lambda\tilde{P}}{\left(1\pm\sqrt{1-\tilde{P}}\right)^{2}+\lambda\tilde{P}}.\label{eq:eta+--1}
\end{equation}
This is the result we presented in {[}Eq. (11){]} of the main text.

\section{\label{sec:Optimal-operation-protocol}Optimal operation protocol
of the heat engine}

Recent studies show that the operation protocols applied to the working
substance will influence the performance of the heat engines \citep{yhmaoptimalcontrol,2020IEGyhma}.
In our model, different operation protocols of the heat engine cycle
will lead to different entropy production $\sigma$, and thus affect
the uniform temperature $\tilde{T}$ of the reservoirs as well as
the power and efficiency of the heat engine. To achieve the boundary
of the trade-off between power and efficiency, in this section, we
aim to find the optimal operation protocol of the heat engine associated
with the minimum entropy production $\sigma_{\mathrm{min}}$.

The minimum entropy production demonstrated in {[}Eq. (5){]} of the
main text is achieved with the following condition
\begin{equation}
\sqrt{\dot{\sigma}}=-\frac{C_{h}\dot{T}_{h}}{\sqrt{L_{22}}}=\frac{\sqrt{\Sigma_{\mathrm{min}}}}{\tau}.\label{eq:minimum entropy production condition}
\end{equation}
In the case of constant $L_{22}$, one has $\Sigma_{\mathrm{min}}=C_{h}^{2}[T_{h}^{[i]}-T_{b}]^{2}/L_{22}=\mathrm{const}$,
which can be simplified as $C_{h}\dot{T}_{h}=const$. Notice that
$\dot{T}_{h}$ represents the change of temperature of the hot reservoir
per cycle, one has 
\begin{equation}
J_{2}\equiv\dot{Q}_{h}=-C_{h}\dot{T}_{h}=\frac{\sqrt{L_{22}\Sigma_{\mathrm{min}}}}{\tau}=\frac{C_{h}[T_{h}^{[i]}-T_{b}]}{\tau},\label{eq:heat absorption of each cycle}
\end{equation}
which means that the heat absorption in each cycle is equal. On the
other hand, in the tight-coupling case with $\vert q\vert=1$, the
heat release to the cold reservoir follows as \citep{izumida2014work}
\begin{equation}
\dot{Q}_{c}=-\frac{T_{c}}{T_{h}}C_{h}\dot{T}_{h}+\frac{T_{c}}{L_{22}}\left(C_{h}\dot{T}_{h}\right)^{2}=\frac{T_{c}}{T_{h}}\frac{C_{h}[T_{h}^{[i]}-T_{b}]}{\tau}+T_{c}\frac{C_{h}^{2}[T_{h}^{[i]}-T_{b}]^{2}}{L_{22}\tau^{2}}.\label{eq:heat release of each cycle}
\end{equation}
Arrive here, we have obtained the heat absorption and the heat release
of each cycle of the entire process.

Before proceeding further, it is worth mentioning that the time scale
$\tau_{e}$, characterizing the temperature change of the reservoirs,
is much greater than the mean cycle time $\tau_{c}$ (microscopic
level), and therefore $T_{h(c)}$ can be regarded as constant in one
cycle. Meanwhile, $\tau_{e}$ is less than the total time $\tau$
(macroscopic level), which implies that in the whole process under
consideration, $T_{h(c)}$ will evolve over time. More precisely,
$T_{h(c)}$ is a step function of $t$ with the intervals being the
time of cycles. By taking the mean cycle time $\tau_{c}$ as the unit
of time, $T_{h(c)}$ is a slowly varying function of $m\equiv t/\tau_{c}$,
($m=1,2,3...M=\tau/\tau_{c}$), i.e., $T_{h(c)}(t)=T_{h(c)}^{(m)}$,
where the new variable $m$ represents the macroscopic dimensionless
time. We stress here that $T_{h(c)}^{(m)}$ represents the temperature
of the reservoir at the beginning of the corresponding finite-time
isothermal process the $m$-th cycle, such that $T_{h(c)}^{(1)}=T_{h(c)}^{[i]}$.
By integrating Eq.(\ref{eq:heat absorption of each cycle}), the temperature
of the hot reservoir in the $m$-th cycle is obtained as
\begin{equation}
T_{h}^{(m)}=T_{h}^{[i]}-\frac{[T_{h}^{[i]}-T_{b}]}{M}\left(m-1\right).\label{eq:T_h}
\end{equation}
Then, combining Eq.(\ref{eq:heat release of each cycle}) and Eq.(\ref{eq:T_h}),
one finds the temperature of the cold reservoir

\begin{equation}
T_{c}^{(m)}=T_{c}^{[i]}\left[1-\frac{[T_{h}^{[i]}-T_{b}]}{MT_{h}^{[i]}}\left(m-1\right)\right]^{-\frac{1}{\gamma}}\exp\left[\frac{C_{h}^{2}[T_{h}^{[i]}-T_{b}]^{2}}{C_{c}L_{22}M\tau}\left(m-1\right)\right].\label{eq:T_c}
\end{equation}

As an example, we specific the working substance of the heat engine
as the ideal gas and assume that the duration of the adiabatic processes
can be ignored \citep{EspositoPRL2010}. In this case, the cycle time
is approximately the sum of the time of two finite-time isothermal
processes. The whole operation of the engine is illustrated in {[}Fig.
4{]} of the main text, where $m$-th cycle contains four thermodynamic
processes A, B, C, and D.

\textbf{($\text{A}$).} In the finite-time isothermal expansion process
of the $m$-th cycle with the duration $t_{h}^{(m)}$, we assume that
the heat flux from high-temperature reservoir to the working substance
satisfies the Newton's law \citep{2020IEGyhma} as
\begin{equation}
q_{h}^{(m)}=\kappa_{h}\left[T_{h}^{(m)}-T_{sh}^{(m)}\right].\label{eq:Newton's law}
\end{equation}
Here, $q_{h}^{(m)}$ is the heat flow, $\kappa_{h}$ is the thermal
conductivity, and $T_{sh}^{(m)}$ is the temperature of the gas which
has been assumed to be unchanged in this sub-process \citep{CA}.
Since the heat absorbed by the working substance from the hot reservoir
in the isothermal expansion process is exactly the total heat released
by the heat reservoir in $m$-th cycle, namely, $q_{h}^{(m)}t_{h}^{(m)}=\dot{Q}_{h}\tau_{c}$,
the temperature of the gas is obtained as 
\begin{equation}
T_{sh}^{(m)}=T_{h}^{(m)}-\frac{C_{h}\left(T_{h}^{[i]}-T_{b}\right)}{M\kappa_{h}t_{h}^{(m)}}.\label{eq:T_sh}
\end{equation}
Alternatively, with the help of the general gas equation $PV=NT_{s}$
($P$ is pressure of the gas, $V$ is volume of the gas, $N$ is the
number of particles and we have taken $k_{B}=1$), the energy conservation
for the gas is expressed as
\begin{equation}
dU^{(m)}=q_{h}^{(m)}dt-\frac{NT_{sh}^{(m)}}{V_{h}^{(m)}}dV_{h}^{(m)}.\label{eq 1st law of hot isothermal process}
\end{equation}
Note that $dU^{(m)}=0$ for the idea gas with constant temperature,
Eq.(\ref{eq 1st law of hot isothermal process}) is further simplified
as 
\begin{equation}
\frac{NT_{sh}^{(m)}}{V}dV=q_{h}^{(m)}dt=\frac{C_{h}\left(T_{h}^{[i]}-T_{b}\right)}{Mt_{h}^{(m)}}dt.\label{eq:differential protocol eqn.}
\end{equation}
Integrating both sides of Eq.(\ref{eq:differential protocol eqn.})
and introducing a new time variable $\tilde{t}\equiv t-\left(m-1\right)\tau_{c}\in\left[0,t_{h}^{(m)}\right]$,
we obtain the optimal protocol of the isothermal expansion as

\begin{equation}
V_{h}^{(m)}\left(\tilde{t}\right)=V_{h,i}^{(m)}\exp\left(\varGamma_{h}^{(m)}\tilde{t}\right),\label{eq:protocol of hot isothermal process}
\end{equation}
where $V_{h,i}^{(m)}$ denotes the initial volume of the gas in the
isothermal expansion, $T_{sh}^{(m)}$ is given in Eq.(\ref{eq:T_sh}),
and the isothermal expansion rate of the $m$-th cycle $\Gamma_{h}^{(m)}$
is defined as
\begin{equation}
\varGamma_{h}^{(m)}\equiv\frac{C_{h}\left(T_{h}^{[i]}-T_{b}\right)}{MNT_{sh}^{(m)}t_{h}^{(m)}}.\label{eq:gammmah}
\end{equation}
 The finial volume of the isothermal expansion follows as $V_{h,f}^{(m)}=V_{h,i}^{(m)}\exp\left(\varGamma_{h}^{(m)}t_{h}^{(m)}\right)$.

\textbf{($\text{B}$).} For adiabatic expansion process of the $m$-th
cycle, according to the adiabatic equation of ideal gas, i.e., $T_{s}V^{\xi-1}=\mathrm{const}.$,
one has 
\begin{equation}
T_{sh}^{(m)}\left[V_{h,f}^{(m)}\right]^{\xi-1}=T_{sc}^{(m)}\left[V_{c,i}^{(m)}\right]^{\xi-1},\label{eq:adiabatic expansion equation}
\end{equation}
where $T_{sc}^{(m)}$ is the constant temperature of the gas during
the isothermal compression, $V_{c,i}^{(m)}$ denotes the initial (final)
volume of the gas in the isothermal compression (adiabatic expansion)
process, and $\xi$ is the heat capacity ratio.

\textbf{($\text{C}$).} In the finite-time isothermal compression
process with the duration $t_{c}^{(m)}$, similar to process \textbf{A},
the exothermic heat flow $q_{c}^{(m)}$ of the $m$-th cycle reads
\begin{equation}
q_{c}^{(m)}=\kappa_{c}\left[T_{sc}^{(m)}-T_{c}^{(m)}\right].\label{eq: New'law in cold isothermal process}
\end{equation}
Note that the heat release from the working substance to the cold
reservoir is exactly the total heat absorbed to the cold reservoir
in $m$-th cycle, namely, $q_{c}^{(m)}t_{c}^{(m)}=\dot{Q}_{c}\tau_{c}$.
Then, with the help of Eq. (\ref{eq:heat release of each cycle}),
the temperature of the gas $T_{sc}^{(m)}$ is obtained as
\begin{equation}
T_{sc}^{(m)}=T_{c}^{(m)}+T_{c}^{(m)}\frac{C_{h}[T_{h}^{[i]}-T_{b}]}{M\sqrt{L_{22}}\kappa_{c}t_{c}^{(m)}}\left(\frac{C_{h}[T_{h}^{[i]}-T_{b}]}{\sqrt{L_{22}}\tau}+\frac{\sqrt{L_{22}}}{T_{h}^{(m)}}\right).\label{eq:T_ch}
\end{equation}
In this process, the energy conservation relation follows as
\begin{equation}
0=C_{s}dT_{sc}^{(m)}=-q_{c}^{(m)}dt-\frac{NT_{sc}^{(m)}}{V_{c}^{(m)}}dV_{c}^{(m)},\label{eq:energy conservation in cold process}
\end{equation}
which determines the optimal protocol of the isothermal compression
as $V_{c}^{(m)}\left(t'\right)=V_{c,i}^{(m)}\exp\left(-\varGamma_{c}^{(m)}t'\right)$.
Here, the isothermal compression rate of the $m$-th cycle $\varGamma_{c}^{(m)}$
is defined as
\begin{equation}
\varGamma_{c}^{(m)}\equiv\frac{T_{c}^{(m)}}{T_{sc}^{(m)}}\frac{C_{h}[T_{h}^{[i]}-T_{b}]}{MN\sqrt{L_{22}}t_{c}^{(m)}}\left(\frac{C_{h}[T_{h}^{[i]}-T_{b}]}{\sqrt{L_{22}}\tau}+\frac{\sqrt{L_{22}}}{T_{h}^{(m)}}\right),\label{eq:Gammac}
\end{equation}
$t'\equiv t-\left[\left(m-1\right)\tau_{c}+t_{h}^{(m)}\right]\,(t'\in\left[0,t_{c}^{(m)}\right])$
is introduced, and $T_{sc}^{(m)}$ is given in Eq.(\ref{eq:T_ch}).
The finial volume of the isothermal compression is calculated as 
\begin{equation}
V_{c,f}^{(m)}=V_{c,i}^{(m)}\exp\left(-\varGamma_{c}^{(m)}t_{c}^{(m)}\right)=V_{h,i}^{(m)}\left[\frac{T_{sh}^{(m)}}{T_{sc}^{(m)}}\right]^{\frac{1}{\xi-1}}\exp\left(\varGamma_{h}^{(m)}t_{h}^{(m)}-\varGamma_{c}^{(m)}t_{c}^{(m)}\right).\label{eq:V_c,f}
\end{equation}

\textbf{($\text{D}$).} For adiabatic expansion process of the $m$-th
cycle, similar to process \textbf{($\text{B}$)}, we have $T_{sc}^{(m)}\left[V_{c,f}^{(m)}\right]^{\xi-1}=T_{sh}^{(m)}\left[V_{h,i}^{(m)}\right]^{\xi-1}$.
In order to make this adiabatic equation consistent with Eq. (\ref{eq:V_c,f}),
the relation $\varGamma_{h}^{(m)}t_{h}^{(m)}=\varGamma_{c}^{(m)}t_{c}^{(m)}$
need to be satisfied. Combining with Eqs. (\ref{eq:gammmah}) and
(\ref{eq:Gammac}), this relation is explicitly obtained as

\begin{equation}
\frac{1}{\kappa_{h}t_{h}^{(m)}}+\frac{1}{\kappa_{c}t_{c}^{(m)}}=\frac{M\left[T_{h}^{(m)}\right]^{2}}{C_{h}[T_{h}^{[i]}-T_{b}]T_{h}^{(m)}+L_{22}\tau},
\end{equation}
which, together with $t_{h}^{(m)}+t_{c}^{(m)}$=$\tau_{c}$, determines
the operation duration of the two finite-time isothermal processes
in each cycle.

The protocols obtained in the above four processes complete the full
operation scheme of the $m$-th cycle. In practice, we fix the initial
value $V_{h,i}^{(m)}$ of the volume of the isothermal expansion in
each cycle, namely, $V_{h,i}^{(m)}=V_{h,i}$. The finial volume $V_{h,f}^{(m)}$
of the isothermal expansion, the initial volume $V_{c,i}^{(m)}$ of
the isothermal compression , and the final volume $V_{c,i}^{(m)}$
of the isothermal compression vary with $m$, and can be explicitly
determined with the full operation protocols.

\bibliographystyle{apsrev}
\addcontentsline{toc}{section}{\refname}\bibliography{../TOFS}